\documentclass[preprint,aps,prl]{revtex4-1}%%,twocolumn
\usepackage{ifpdf}
\usepackage{times}
\usepackage{amsmath}
\usepackage{amssymb}
\usepackage{units}
\usepackage{stmaryrd} %% "// sign" = \sslash
\usepackage[frenchb]{babel}
\usepackage[T1]{fontenc}
\usepackage[applemac]{inputenc} 	
\usepackage{graphicx}
\usepackage{dcolumn}% Align table columns on decimal point
\usepackage{bm}% bold math
\usepackage{color}

\def\be{\begin{equation}}
\def\ee{\end{equation}}
\def\bea{\begin{eqnarray}}
\def\eea{\end{eqnarray}}

\renewcommand{\vec}[1]{\boldsymbol{#1}}

\begin{document}
\preprint{0}

\title{Confinement of superconducting fluctuations due to emergent electronic inhomogeneities}

\author{C. Carbillet$^{1}$, S. Caprara$^{2,3}$, M. Grilli$^{2,3}$, C. Brun$^{1}$, T. Cren$^{1}$, F. Debontridder$^{1}$, 
B. Vignolle$^4$, W.\,Tabis$^4$, D. Demaille$^1$, L. Largeau$^5$, K. Ilin$^6$, M. Siegel$^6$, D. Roditchev$^{1,7}$, 
and B. Leridon$^{7 *}$}

\affiliation{$^1$UMR 7588, Institut des nanoscsiences de Paris, Sorbonne Universit\'{e}s, UPMC, CNRS, F-75005  
Paris, France\\
$^2$ Department of Physics, Universita La Sapienza, Piazzale A. Moro 5, I-00185 Rome, Italy\\ $^3$ ISC-CNR, 
via dei Taurini  19, I-00185 Roma, Italy\\ $^4$Laboratoire National des Champs Magn\'{e}tiques Intenses, UPR 3228, 
(CNRS-INSA-UJF-UPS), F-31400 Toulouse, France\\ $^5$Laboratoire de Photonique et Nanostructures UPR20/CNRS, 
Route de Nozay, F-91460 Marcoussis, France\\ $^6$ Institute of Micro- und Nano-electronic Systems, Karlsruhe Institute 
of Technology, Hertzstrasse 16, D-76187 Karlsruhe, Germany\\ $^7$  
ESPCI ParisTech, PSL Research University,  CNRS, Sorbonne Universités, UPMC Univ. Paris 6, Laboratoire de Physique et d'Etude des Matériaux (LPEM), 10 rue Vauquelin, F-75231 Paris Cedex 5, France \\* corresponding author: brigitte.leridon@espci.fr}
 
\begin{abstract}
The microscopic nature of an insulating state in the vicinity of a superconducting state, in the presence of 
disorder, is a hotly debated question. While the simplest scenario proposes that Coulomb interactions destroy 
the Cooper pairs at the transition, leading to localization of single electrons, an alternate possibility 
supported by experimental observations suggests that Cooper pairs instead directly localize. The question of 
the homogeneity, granularity, or possibly glassiness of the material on the verge of this transition is intimately 
related to this fundamental issue. Here, by combining macroscopic and nano-scale studies of superconducting 
ultrathin NbN films, we reveal nanoscopic electronic inhomogeneities that emerge when the film thickness is reduced. 
In addition, while thicker films display a purely two-dimensional behaviour in the superconducting fluctuations, 
we demonstrate a zero-dimensional regime for the thinner samples precisely on the scale of the inhomogeneities. 
Such behavior is somehow intermediate between the Fermi and Bose insulator paradigms and calls for further investigation to understand the way Cooper pairs continuously evolve from a bound state of fermionic objects into 
localized bosonic entities.

\end{abstract}

\maketitle

Understanding the microscopic processes occurring at the superconductor-insulator transition (SIT) in 
ultrathin films is important not only for fundamental reasons, but also for applicative purposes
\cite{Goltsman:2001ea,Hofherr:2010iy}. The microstructural properties are known to play a key role and 
the samples can then be divided into two groups \cite{Goldman:1998jy}: granular thin films and homogeneously 
disordered thin films. For the former, the SIT is driven by the competition between the intergrain Josephson 
coupling, favoring the delocalization of pairs, and the Coulomb charging energy of the grains, which renders 
charge fluctuations energetically expensive \cite{Beloborodov:2007hq}. In the case of nominally homogeneously 
disordered  films however, which are the object of this Article, several scenarios have been proposed. On the one 
hand, what is often referred to as the "fermionic" scenario proposes that the mechanism that drives the transition 
is the reduced screening of the Coulomb repulsion with increasing disorder, weakening pairing and reducing the 
critical temperature $T_C$ \cite{FinkelShtein:1987tl}, as observed in \cite{VallesJr:1992va}. In this case, the insulating state is composed of localized 
fermions and, in particular, standard paraconductive fluctuations are expected due to Gaussian-distributed 
short-lived Cooper pairs. On the other hand, pairs may survive the SIT in a "bosonic" scenario, in which the gap 
persists above $T_C$ despite the loss of phase coherence. In this framework, the bosonic pairs either localize 
because disorder-enhanced Coulomb interactions destroy their phase-coherent motion at large scales 
\cite{Fisher:1990zza,Fisher:1990zz} or disorder itself can blur the pair phase coherence without any relevant role 
of the Coulomb repulsion \cite{castellani:2012,Feigelman:2007bq,Feigelman:2010hp,Ioffe:2010dp}. On the experimental side, it has been shown through careful study of Little-Parks oscillations, that either fermionic  \cite{Hollen:2013ht} or bosonic \cite{Stewart:2007uz}\cite{Kopnov:2012bx} transitions may occur. 
In the latter case, it was also proposed that the superconducting state is characterized by an emergent 
disordered glassy phase \cite{Ioffe:2010dp} with filamentary superconducting currents \cite{castellani:2012}.  
An anomalous distribution of the superconducting order parameter  was proposed by theorists 
\cite{Ioffe:2010dp,Lemarie:2013bk}, and observed experimentally \cite{Sacepe:2008jx,Sacepe:2011jm}. 
A numerical approach to uniformly disordered superconductors \cite{Bouadim:2011hx} has also suggested that there is 
a continuous evolution \cite{Trivedi:2012cj} from the weak disorder limit, were the system has a rather 
homogeneous fermionic character, to the strong disorder limit, where marked inhomogeneities appear in the 
superconducting order parameter, with an emergent bosonic nature characterized by a single-particle gap persisting 
on the insulating side. A great deal of experimental activity has been devoted to the more disordered part of the 
SIT \cite{Sacepe:2008jx,Kamlapure:2013kh} while the intermediate region where fermionic Cooper pairs begin to evolve 
into bosonic pairs has not been extensively accessed. The aim of this work is precisely to fill this gap by reporting experiments which shed light on how Cooper pairs evolve with increasing disorder, giving rise to incipient 
inhomogeneous bosonic features. More specifically, we present here a study on a set of ultrathin NbN films that are nominally homogeneous, but where electronic inhomogeneities and pseudogap emerge as the thickness is reduced, together 
with experimental indications in favor of fermionic mechanisms. Indeed, while the thickest films ($d >2.2$\,nm, 
$T_C \geq 0.3\, T_C^{bulk}$) are found to be rather homogeneous \cite{Noat:2013bu} with two-dimensional (2D) 
Aslamazov-Larkin (AL) superconducting fluctuations, the thinnest samples offer a more complex behavior characterized 
by inhomogeneous superconductivity, indication for a pseudogap above $T_C$ (a seemingly "bosonic" feature), 
in agreement with the literature \cite{Sacepe:2008jx,Sacepe:2010gt,Sacepe:2011jm} and by the establishment of a 
zero-dimensional (0D) regime of gaussian superconducting fluctuations (a fermion-like hallmark).
We show that for films below this threshold film thickness and critical temperature, the superconducting fluctuations measured by transport above $T_C$ behave in agreement with a formal 0D limit of the AL theory of paraconductivity, in 
a substantial range of reduced temperature $\epsilon\equiv\log(T/T_C)$. This indicates that these fluctuations are 
still conventional and seemingly BCS-like, but somehow confined in a "supergrain" over a lengthscale $l$. 
The superconducting inhomogeneities as evidenced by scanning tunneling spectroscopy at low temperature correspond 
to electronic domains of typical size $L_i/2$ precisely of the order of $ l \sim 50\,nm $ i.e., 
much larger than any definite structural scale of the system.
%while it  is of the same order of, but totally uncorrelated from,the inhomogeneous smooth landscape found in the sample topography.  
The paradoxical presence of AL fluctuations together with indications for bosonic features such as the pseudogap, 
well established in these systems \cite{Sacepe:2008jx,Sacepe:2010gt,Sacepe:2011jm}, is one of the most intriguing 
results of this work. We suggest that these two features can indeed be reconciled if the pseudogap in our 
system has a fluctuational origin \cite{Abrahams:1970tn}. We propose that the pronounced amplitude of the 
pseudogap observed in the ultrathin films that are the object of our investigation arises from a slowing down in the 
diffusion of the fluctuating Cooper pairs, which exhibit a tendency to localize on the typical inhomogeneity scale 
$L_i$ and, at the same time, give rise to the 0D AL behavior.

%%%%%%%%%%%%%%%%%%%%%%%%%%%%%%%%
\vspace{0.5 truecm}
{\bf Results}
%%%%%%%%%%%%%%%%%%%%%%%%%%%%%%%%

Our samples consist of ultrathin NbN films grown ex situ on sapphire substrates. Details of the fabrication process 
may be found in \cite{Semenov:2009em}. (See also the Appendix). The different 
samples studied together with their thickness, critical temperature, and resistance per square at room temperature 
may be found in Table 1 of the Appendix.
%------------------------------------
\begin{figure*}
  \centering
 \includegraphics[width = \textwidth]{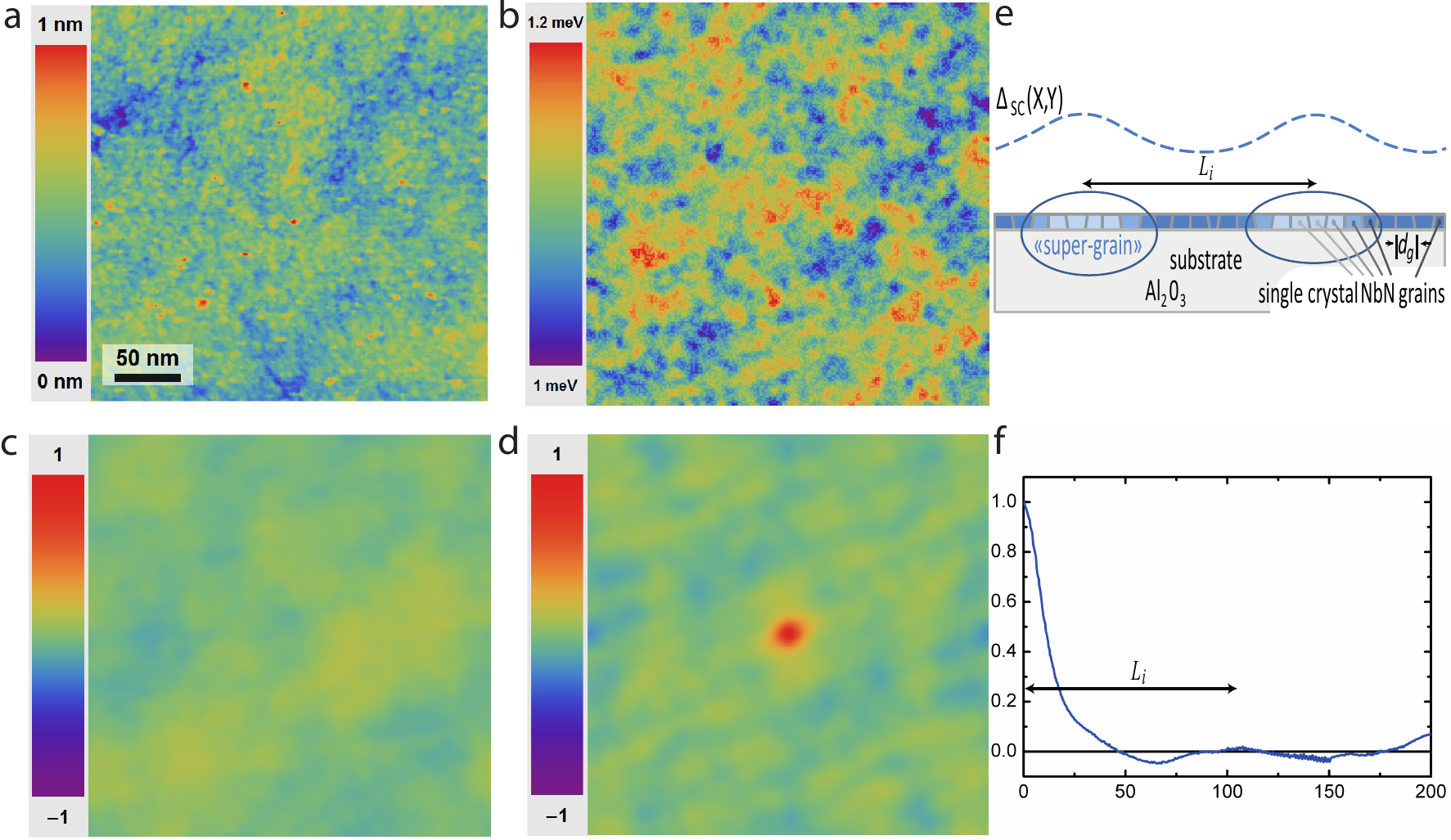}
\caption{Local inhomogeneities of the superconducting state. (a) Topographic map of the NbN area under study for 
sample $X_0$ ($T_C= 3.8$\,K). (b) Corresponding map displaying the superconducting gap inhomogeneities at 300\,mK. 
Gap inhomogeneities  appear on a much larger scale than the size of the nanocrystals ($d_g \sim 2-5$\,nm) constituting 
the NbN films and seen in the topographic map.(c ) Cross-correlation map between the topographic and spectroscopic 
maps revealing the absence of correlation. (d) Autocorrelation map of the gap map (b). (e) Schematics of the 
relevant length scales. (f) Radial profile extracted from the autocorrelation map (d). The correlation length 
$L_i$ is defined as the abscissa of the first peak away from the center and so is about 100\,nm in the present sample. 
The central peak width, about $L_i/2 \sim 50$\,nm gives an estimate of the typical domain size.}
  \label{STS}
\end{figure*}
%------------------------------------

Transmission Electron Microscopy (TEM) measurements of our NbN films (see Fig.5 in the Appendix) were able to 
show that the films are smooth and well crystallized, and can be viewed as a closely packed assembly of 
nanocrystallites of different orientations. These contiguous nanocrystals have typical lateral dimensions $d_g$ of 
about 2$-$5\,nm. The characteristic film structure is also reflected in topographic scanning tunneling microscopy 
images, as the one presented in Fig.~\ref{STS}a for a nominally 2.14\,nm-thick sample $X_0$ ($T_C$=3.8\,K). 
The film surface is very flat and the observed nanoscale structures correlate well with the nanocrystals revealed 
by TEM. At the same time the landscape of the sample also displays smooth 
inhomogeneities on a larger scale of a few tens of nanometers.

{\em Probing the inhomogeneities with STM and STS} ---
In order to get insight into the superconducting (in)homogeneity of these ultrathin films, we performed scanning 
tunneling microscopy (STM) and spectroscopy (STS) experiments. Typical results are displayed in Fig.~\ref{STS}. 
Fig.~\ref{STS}(a) shows a topographic map of the area under study for sample $X_0$ ($T_C=3.8$\,K). In Fig.~\ref{STS}(b) 
we report the extracted superconducting gap map $\Delta(X,Y)$ at 300\,mK i.e. well inside the superconducting state 
for the area corresponding to the topographic map of Fig.~\ref{STS}(a). $\Delta(X,Y)$ is defined from the $dI/dV$ 
spectra as the peak-to-peak energy (see Fig. 6 in the Appendix for the corresponding representative $dI/dV$ spectra).
In agreement with previous measurements in similar systems \cite{Sacepe:2008jx,Kamlapure:2013kh},
we observe spatial variations of the gap (see Fig.~\ref{STS}(b)) and of the integrated in-gap conductance (see Appendix). 
However, we reveal here for the first time that \textit{these gap variations are not correlated to the topography of the surface}. 
Indeed, the cross-correlation map between (a) and (b) reveals the absence of correlation 
[see Fig.~\ref{STS}(c)].

In Fig.~\ref{STS}(d) we report the autocorrelation map of the gap map shown in Fig.~\ref{STS}(b). It allows us to 
extract both the typical size of domains of constant gap values, hereafter called  supergrains, and the typical 
distance $L_i$ between the centers of such adjacent supergrains. The radial profile extracted from the center of 
the autocorrelation map is shown in Fig.~\ref{STS}(f), featuring a correlation length $L_{i}$ of about 100\,nm and 
a typical domain size of about $L_i/2$. This is more than ten times larger than the size of the microstructural grains 
$d_g\simeq2-5$\,nm (see Fig. \ref{STS}(e) depicting the relevant lengths scales). Above the global critical 
temperature $T_C$ these inhomogeneities persist, as well as a gap-like feature called pseudogap. In addition, 
our measurements reveal that the energy integrated in-gap conductance maps at 300\,mK and at 4.2\,K, performed on the 
same area of Fig.~\ref{STS}(a) are correlated (see Fig. 6 in the Appendix). We deduce 
that at 4.2\,K $L_{i}$ remains roughly the same, whilst the spectroscopic contrast between the regions changes 
with temperature.

By contrast, similar spectroscopic studies carried out on thicker samples 
($d\geq 2.3$\,nm, $T_C \geq 0.3\,T_C^{bulk}$) revealed a much more homogeneous superconducting phase 
\cite{Noat:2013bu}. Therefore, some short-scale inhomogeneity in the superconductive properties emerges for the 
thinner samples at a scale $L_i$ while the system is structurally homogeneous over the same length scale. This is 
consistent with predictions from Monte-Carlo simulations \cite{Bouadim:2011hx}.

\begin{figure}
\centering
 \includegraphics[width = \textwidth]{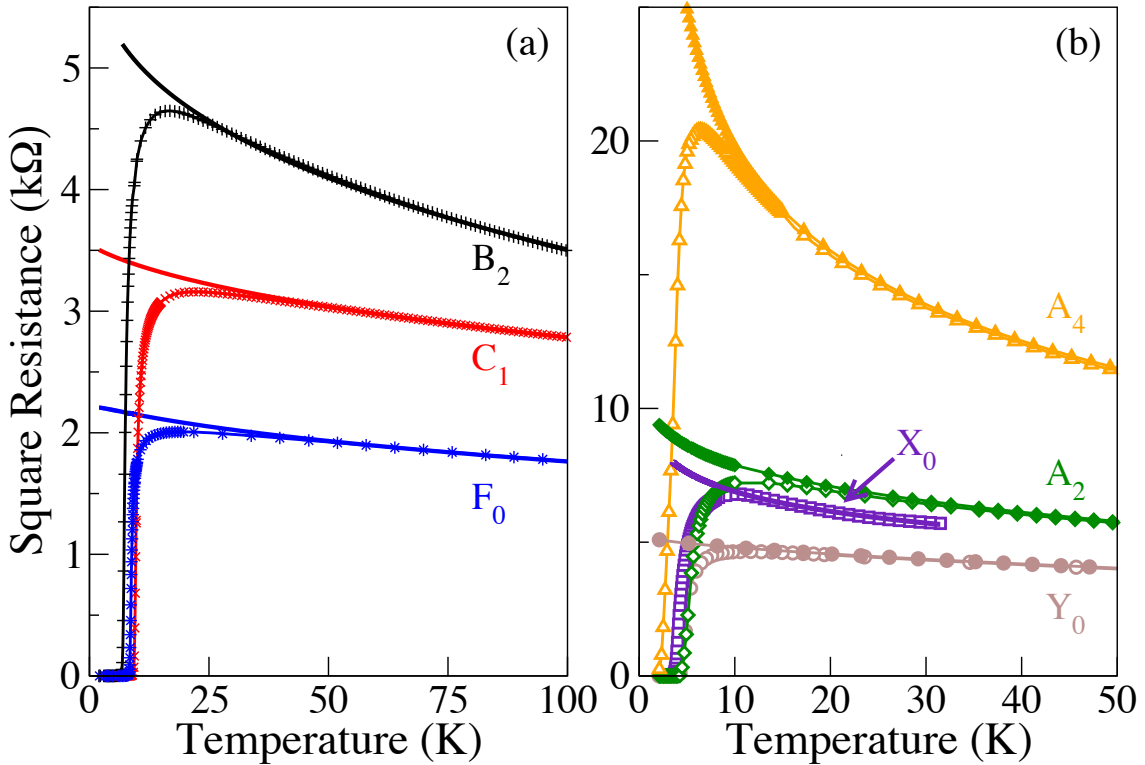}
  \caption{(a) Square resistance under 0\,T as a function of temperature for the thicker samples 
  $B_2$ ($T_C=7.1$\,K, $+$), $C_1$ ($T_C=9.4$\,K, $\times$), $F_0$  ($T_C=9.0$\,K, $*$), with the corresponding 
  extrapolated normal state resistance (solid lines, see text). (b) Square resistance under 0\,T as a function of temperature for the thinner   
  samples $Y_0$ ($T_C=4.3$\,K, open circles), $X_0$ ($T_C=3.8$\,K, open squares), $A_2$ ($T_C=4.5$\,K, open diamonds),
  and $A_4$ (aged $A_2$, $T_C=2.4$\,K, open triangles). The square resistances under a perpendicular magnetic field 
  of 14\,T are reported with the corresponding full symbols. The solid lines are for the extrapolated normal state resistances. }
  \label{resnor}
\end{figure}

{\em Analysis of the paraconductivity} ---
In order to probe the influence of these nanoscale inhomogeneities on the superconducting thermal fluctuations, 
we performed transport measurements in the vicinity of $T_C$ and extracted the paraconductance  per square
$\Delta\sigma (T)= \sigma (T) -\sigma_N(T)$, i.e., the excess conductance per square due to superconducting fluctuations 
in the normal state. Here, $\sigma(T)$ is the square conductance measured under zero magnetic field and $\sigma_N(T)$ is
the normal state square conductance. The resistance per square  is displayed on Fig.\ref{resnor} for the different samples, together with the extrapolated [(a) solid lines] or measured [(b) full symbols] resistance of the normal state. 
(See the methods Section for details on the determination of the normal state.)

\begin{figure}
\centering
 \includegraphics[width = \textwidth]{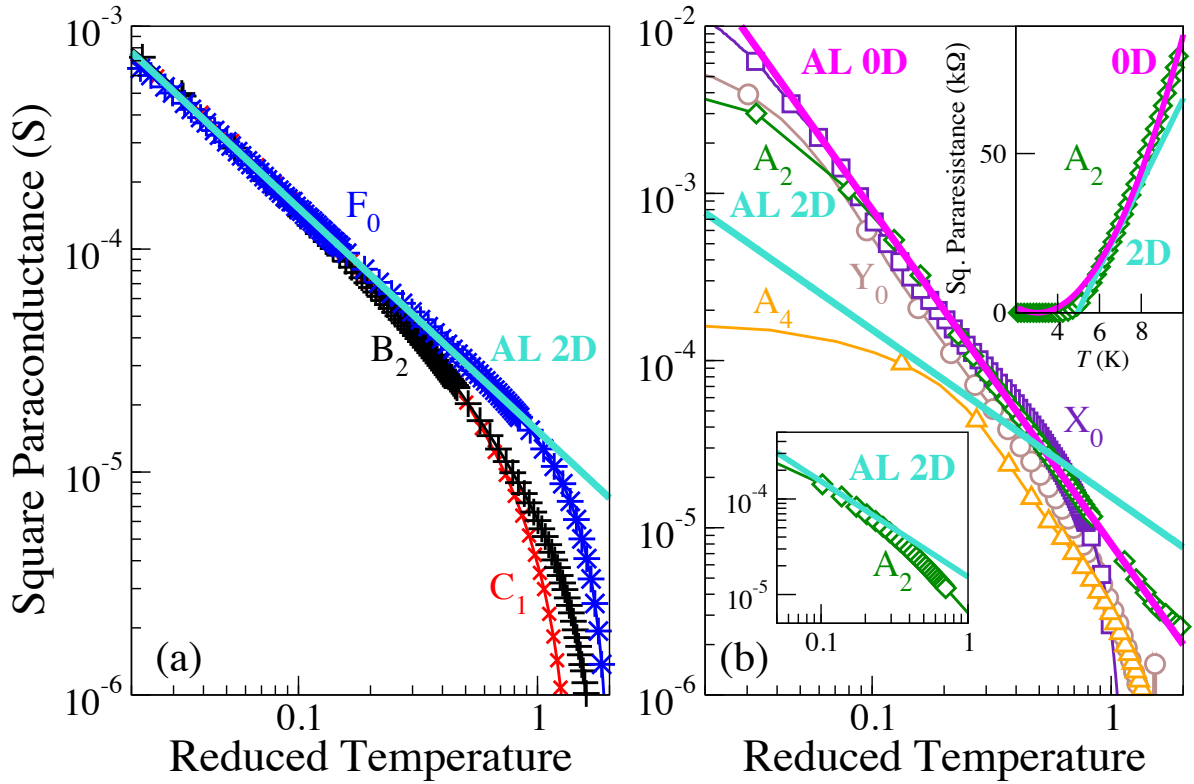}
  \caption{(a) Extracted square paraconductance for the thicker samples $B_2$ ($T_C=7.1$\,K, $+$), $C_1$  
  ($T_C=9.4$\,K, $\times$), and $F_0$ ($T_C=9.0$\,K, $*$) as a function of the reduced temperature 
  $\epsilon\equiv \ln(T/T_C)$. The agreement with the Aslamasov-Larkin prediction for a 2D system (purple solid line) 
  is excellent, without any adjustable parameter. (b) Extracted square paraconductance for the thinner samples 
  $Y_0$ ($T_C=4.3$\,K, open circles), $X_0$ ($T_C=3.8K$, open squares), $A_2$ ($T_C=4.5$\,K, open diamonds), 
  $A_4$ ($T_C= 2.4K$, open triangles). The maroon solid line corresponds to $\Delta \sigma= 0.03e^2/(\hbar \epsilon^2)$.
  The expected AL 2D square paraconductance is also shown (purple solid line). The upper inset shows the pararesistance
   $\Delta \sigma^{-1}$ as a function of temperature for the sample $A_2$, emphasizing the 2D [$\propto (T-T_C)$]
   and 0D [$\propto (T-T_C)^2$] behaviours, the crossover form one to the other, as well as the different 
   fluctuative critical temperatures in the two regimes ($T_C^{2D}=4.9$\,K, $T_C^{0D}=4.4$\,K). The lower inset shows 
   the crossover to the 2D AL square paraconductance for the sample $A_2$ as a function of the reduced temperature. (In this case $\epsilon=ln(T/T_C^{2D})$ was used.)}
  \label{AL}
\end{figure}

In Fig.\,\ref{AL}(a) the variation of $\Delta \sigma$ with the reduced temperature 
$\epsilon\equiv \ln(T/T_C)$ is shown for samples $B_2$, $C_1$, and $F_0$: the observed critical exponent $-1$ is 
consistent with the Aslamasov-Larkin prediction for 2D systems (AL 2D) \cite{Aslamasov:1968vx} 
$\Delta \sigma=e^2/(16\hbar \epsilon)$ in the range $0.02\leqslant \epsilon \leqslant 0.2-0.9$, as it was 
reported previously for these films \cite{Semenov:2009em}. Remarkably, the \textit{extracted experimental 
AL 2D prefactor matches the theoretical one}, without any adjustable parameter. This suggests that the fluctuations 
in this case are BCS-like, and that Maki-Thomson (MT) fluctuations \cite{Maki:1968vo, Thompson:1970ue} or 
density-of-state (DOS) corrections\cite{Beloborodov:1999va,Beloborodov:2000vw,Beloborodov:2007hq} are absent or 
negligible. 

Proceeding in a similar way, we extracted the paraconductivity of the thinner samples $A_2$, $A_4$ (aged $A_2$), 
$X_0$ and $Y_0$ (shown as open symbols in Fig.\,\ref{AL}(b). The AL 2D prediction is also displayed in the same 
Figure (purple solid line). The experimental paraconductance is found to deviate strongly from the AL 2D behavior 
over a significant temperature range, even when using $T_C$ as a free parameter, and to exhibit over
a substantial range of reduced temperature a very specific law, $\Delta\sigma \sim \epsilon^{-2}$, corresponding to 
0D fluctuations, previously observed in granular materials 
\cite{Kirtley:1974tb, Deutscher:1974wn,Civiak:1976uh,WOLF:1977wl}.
An empirical fitting function, $\Delta \sigma= 0.03e^2/(\hbar\epsilon^2)$ for $A_2$ and $X_0$ is plotted
for comparison (maroon solid line). Similar behaviors, i.e., $\Delta \sigma= 0.02e^2/(\hbar\epsilon^2)$
for $Y_0$, and $\Delta \sigma= 0.015e^2/(\hbar\epsilon^2)$ for $A_4$, were found to hold for the other samples.
The paraconductance data, e.g., for sample $A_2$ also suggests a 0D-2D crossover [see also the lower inset
in Fig.\,\ref{AL}(b)] as previously observed in 
\cite{Belevtsev:1983wd} and discussed in \cite{Deutscher:1974wn}. We point out that the fluctuational critical
temperature may differ in the two regimes, so that the crossover is more evident when the pararesistivity 
$\Delta\sigma^{-1}$ is plotted as a function of $T$, without making any choice for $T_C$, rather than
$\epsilon$, which depends on $T_C$ [see the upper inset in Fig.\,\ref{AL}(b)].

Concerning the absence of MT terms, we stress out that, with pair breaking arising only from 
electron-electron interactions, MT paraconductance is less singular than the AL term in 2D \cite{Reizer:1992wr}, 
and even less so for 0D AL. On the other hand, the presence of a sizable pseudogap suggests that DOS 
corrections should be present. DOS corrections, however, lead to a decrease of the paraconductance. In our case, 
instead, the paraconductance in the pseudogap regime is found to be even more singular with $\epsilon^{-2}$ 
dependence that can not be explained by DOS contribution. This clearly indicates that DOS corrections, although 
expected, are immaterial in this case.

\begin{figure}
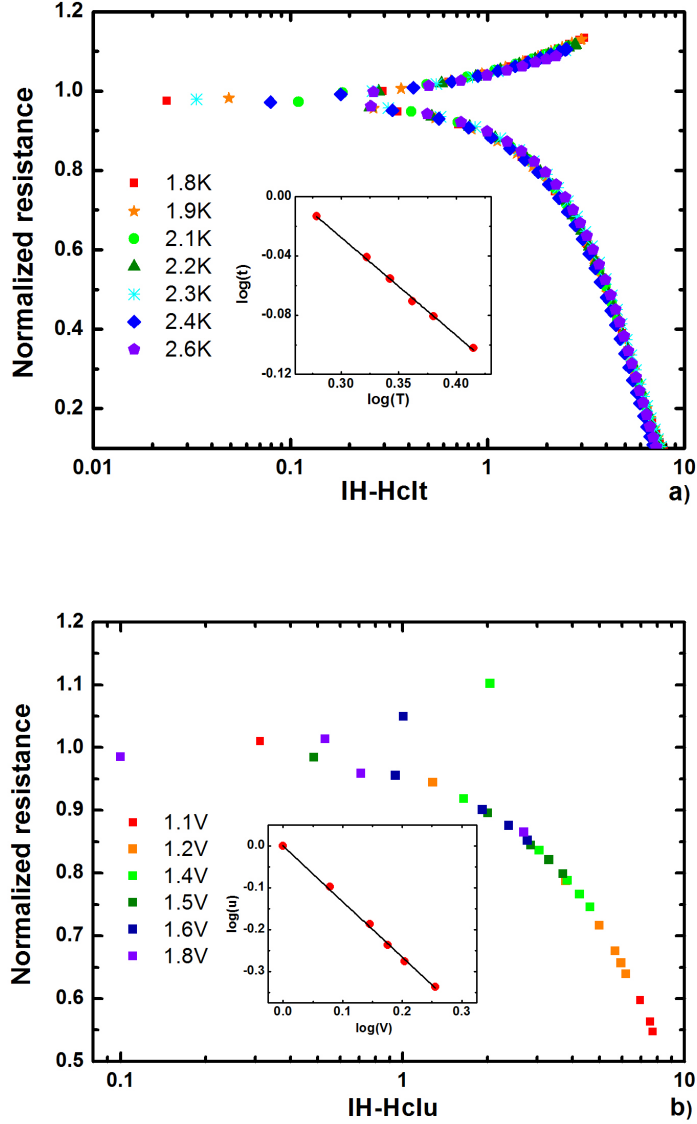

  \centering
  \includegraphics[width = 0.7\textwidth]{Fig4a.pdf}
   \includegraphics[width = 0.7\textwidth]{Fig4b.pdf}
  \caption{(a) Normalized resistance $R/R_C$ of sample $A_2$ as a function of the scaling variable $|H-H_C|t$ for 
  different temperatures. $H_C=10.7$\,T and $R_C=11$\,k$\Omega$ at the crossing point. $t\equiv T^{-1/\nu z}$ was 
  adjusted in order to obtain the best collapse of the data. Inset: Log-log plot of the parameter $t$ vs temperature, 
  used to determine the value of $\nu z=2/3$. (b) Normalized resistance $R/R_C$ of sample $A_2$ as a function of 
  the scaling variable $|H-H_C|u$ for different electrical fields, measured at $T=$1.9\,K. $u\equiv E^{-1/\nu(z+1)}$ 
  was adjusted in order to obtain the best collapse of the data. Inset: Log-log plot of the parameter $u$ vs. voltage 
  used to determine the value of $\nu(z+1)=4/3$.}
  \label{QSIT}
\end{figure}

{\em Study of the magnetic field driven transition} ---
The above study near $T_C$ was complemented by the analysis of the transport properties of the thinnest samples at 
the transition to the normal state driven by magnetic field, well below $T_C$. We found that the curves $R(H)$ cross 
at a specific point $\left\{R_C;H_C\right\}$ (see also Fig.\,8 in the Appendix for further 
details). Following Ref. \onlinecite{Sondhi:1997wz}, we analyzed the curves in the vicinity of this point 
and evidenced a scaling behavior of the type $R/R_C=f(|H-H_C|T^{-\alpha})$ with the critical exponent 
$\alpha\approx 3/2$ (See the data in Fig.\,\ref{QSIT}(a) for sample $A_2$). The occurrence of such scaling behavior, where $T$ is the only relevant scale, marks the existence of a quantum critical point 
(QCP) at zero temperature and for $H=H_C$. Consequently, the exponent $\alpha$ can be expressed as 
$\alpha=1/(\nu z)$, where $\nu$ is the exponent that rules the variation of the spatial correlation length 
$\xi\sim \vert H-H_c \vert^{-\nu}$ and $z$ is the dynamical critical exponent $\xi^z\sim 1/T$.

In order to extract the values of $\nu$ and $z$ from $\alpha\approx 3/2$, similar measurements were performed at a 
fixed temperature, $T=$1.9\,K, for different values of the electric field, i.e., of the bias voltage across the sample 
[see the data for sample $A_2$ in Fig.~\ref{QSIT}(b)]. Here again, the $R(H)$ curves exhibit a common crossing 
point corresponding to the QCP (see Fig.~9 in the Appendix).
The scaling analysis in the vicinity of this point, with a scaling function of the form
$R/R_C=g(|H-H_C|E^{-\beta})$, yielded $\beta\approx 3/4$. Following the analysis in
Refs.\,\onlinecite{Sondhi:1997wz,Yazdani:1995ut,Markovic:1999vy}, we express $\beta$ as $\beta=1/\nu(z+1)$. 
The two independent determinations of $\nu z$ and $\nu (z+1)$ allow us to establish $\nu = 2/3$ and $z=1$. The latter 
is precisely the result expected, e.g., in systems with (weakly screened) long-range Coulomb interactions, while the 
former is consistent with a 3D classical XY model or 2D quantum XY model (with z=1).
A similar analysis, performed on sample $F_0$ under high pulsed magnetic field allowed to evidence a plateau between 1.5K and 8K for $H_C=18.6K$ and to extract a product of critical exponents $\nu z \sim 2/3$.

%%%%%%%%%%%%%%%%%%%%%%%
\vspace{0.5 truecm}
{\bf Discussion}
%%%%%%%%%%%%%%%%%%%%%%%

As we report in the Appendix, the coherence length exponent that leads to the observation 
of an anomalous power law $\Delta\sigma \sim \epsilon^{-2}$, is consistent with a {\it formal} calculation of 
AL fluctuations in a 0D system. $\Delta\sigma$ is converted into the measured paraconductance per square by means of 
a suitable length scale $l$, which represents the size of the 0D fluctuating domains, in the plane parallel the film, 
yielding $\delta \sigma_{D=0}=(\frac{\xi_0}{l})^2\frac{\pi e^2}{4\hbar\epsilon^2}$.
Deducing for $\xi_0$ a value of 5.5\,$\pm$\,0.5\,nm from the value of $H_C$ at the QCP, in agreement with extrapolated 
values in \cite{Semenov:2009em}, it is possible to extract $l$ from the paraconductivity data. One finds, e.g., 
$l=28$\,nm for samples $A_2$ ($T_C=4.5$\,K) and $X_0$ ($T_C=3.8$\,K), $l=35$\,nm for sample $Y_0$ ($T_C=4.3$\,K), 
and $l=40$\,nm for sample $A_4$. This length $l$ is in good quantitative agreement with the typical domain 
size $L_i/2 \sim 50\,nm$ extracted from STS 
data at 300\,mK and at 4.2\,K for sample $X_0$ ($T_C=3.8$\,K). This means that the length $L_i/2\sim l$ 
\textit{and not the real grain size $d_g \ll l$} sets the scale for the 0D fluctuating domains, until $\xi$
becomes larger than $L_i$.  Further theoretical investigation is necessary to understand the physical meaning of 
the presence of these 0D domains, that behave like smooth "supergrains". The observed behavior could actually 
be interpreted in terms of slower diffusion of the Cooper pairs thus increasing their lifetime on length scales 
smaller than $l$. By contrast, a standard diffusive behavior is recovered at longer times and larger distances giving 
way to standard 2D behavior eventually ruling the transition (see the Appendix for more details).

To our knowledge, the only previous evidence in the literature of a 0D fluctuations regime in transport is 
for nominally granular or filamentary systems \cite{Kirtley:1974tb,Civiak:1976uh,WOLF:1977wl}. The novelty lies 
here in the observation of such behavior in a compound where the inhomogeneity arises in a ``mild'' way: The films 
are far from granularity because the 0D behavior does not occur on the small scale of the crystallites, but rather 
on the larger typical scale $l$, comparable to the correlation length $L_i/2$ inferred from STS. The emergent 
(as opposed to the structural) character of the 0D regions is also suggested by the lack of any correlation 
[see Fig.~\ref{STS}(c)] between the inhomogeneous domains observed  with STS [Fig.~\ref{STS}(b)] and the 
large-scale structural disorder observed in the topography of the crystallite ensemble [Fig.~\ref{STS}(a)].

For the superconducting transition to be probed in transport, the 0D fluctuations regime has
to finally evolve to higher-dimensional behavior. Close enough to $T_C$ a crossover to 2D behavior must (and does)
occur, when the superconducting coherence length becomes larger than $L_i$ and enables to couple
different 0D domains, following a scenario analogous to the Lawrence-Doniach description for lamellar materials 
\cite{Lawrence:1971,Deutscher:1974wn}.  As a matter of fact, the 0D-2D crossover is visible for sample $X_0$ 
($d=2.14$\,nm), as well as for the sample $A_2$ ($d=2.16$\,nm). Please see the insets of Fig. \ref{AL}.
However, we propose here a different physical explanation for this crossover.
We start with an expression of the AL paraconductivity in $D$ dimensions (see details in  the Appendix):
\be
\delta\sigma_D(\epsilon)=\frac{\pi e^2}{4\hbar D}\int_{0}^{\infty} \mathrm d\gamma \,\frac{\mathcal N_D(\gamma)}{(\epsilon+\gamma)^3}  \label{eq1}
\ee
with a suitable "density of states" (weighted with current vertices) $\mathcal N_D(\gamma)$. Setting $D=2$, the standard AL result
is found for
$\mathcal N_2(\gamma)=\gamma/\pi$, corresponding to diffusion of fluctuating Cooper pairs in two dimensions. However, if
\be
\mathcal N_2(\gamma)=\left\{ {\gamma/\pi,~~~~~~\mathrm{for}~\gamma\le\bar\gamma,\atop  \bar\gamma/\pi,~~~~~~\mathrm{for}~\gamma>\bar\gamma,}\right. 
\label{eq2}
\ee
corresponding to a slowing-down of the diffusion of fluctuating Cooper pairs above a threshold $\bar\gamma$, a 0D behavior
\be
\delta\sigma\approx\frac{e^2\bar\gamma}{16\hbar\epsilon^2}  \label{eq3}
\ee
is found for $\epsilon\gg \bar\gamma$. A comparison with the formal extrapolation of AL fluctuations to $D=0$ yields
$\bar\gamma=4\pi(\xi_0/l)^2$. For $\epsilon\ll \bar\gamma$, the standard 2D AL paraconductivity is recovered.

Finally, at even lower temperature (below the 2D AL transition temperature), the system
might be governed by percolation physics or, alternately, by Berezinski-Kosterlitz-Thouless behavior.
This behavior, if any, should occur on a very restricted range of temperature,  because our measurements 
always display the paraconductivity of standard gaussian fluctuations.
If the bosonic scenario was applicable to our films, the transition should be ruled by dephasing of the pairs 
and paraconductivity should mirror the characteristics of the vortex fluctuations in the BKT transition 
\cite{Halperin:1979to}. So, at first sight, our findings seem at odds with the existence of a pseudogap
on the normal side of the transition as is well established in the literature and can be seen in the conductance 
maps and $dI/dV$ spectra in the supplementary material, and which is usually ascribed to the strong localization 
of Cooper pairs \cite{Sacepe:2010gt,Sacepe:2011jm}. Our work suggests a different explanation. The diffusion 
slowing down of the fluctuating Cooper pair in the "supergrains" increases their lifetime. This might occur either 
because they locally find a more suitable environment, like, e.g., a locally higher $T_C$, or, conversely, because 
they display an increased tendency to localize for those wavevectors corresponding to the largest inhomogeneities. 
This second possibility seems more likely because our fits suggest that the 0D critical temperature (i.e. the ``local'' critical temperature) is slightly lower than the 2D large-scale $T_C$. The incipient localization and 
diffusion slowing down of the Cooper pairs seems to have a sizable effect on depressing the density of state at 
the Fermi level, thus suggesting the possibility of a substantial \textit{fluctuational} pseudogap 
\cite{Abrahams:1970tn} \cite{Beloborodov:1999va}.

The magnetic-field driven transition at low temperature can be interpreted as magnetic field-induced dephasing of 
the 0D supergrains thereby accounting for the critical exponents $\nu=2/3$ of an XY model in 2+1 
dimensions, the additional "1" coming from the dynamical critical index $z=1$. The value $z=1$ is pertinent, e.g., 
to systems with (weakly screened) long-range Coulomb interactions \cite{Herbut:2001hn}. It is also consistent 
with numerical calculations based on a Boson-Hubbard model \cite{Kisker:1997un}. Similar values for $\nu$ were 
observed in Bi \cite{Markovic:1999vy} or NbSi \cite{Aubin:2006ju} amorphous thin films, whereas a large number of 
studies of the SIT point towards a different universality class with $\nu=4/3$ 
\cite{Hebard:1990up,Yazdani:1995ut,Kapitulnik:2001cu,Steiner:2005ks}, a critical exponent consistent with classical
percolation. Our findings are therefore in agreement with a picture of phase fluctuating 0D supergrains coupled 
\textit{\`a la} Josephson, \textit{for the magnetically driven transition}.

The question arises of the origin of the electronic inhomogeneities. Indeed, while the structural small 
scale inhomogeneity associated to the structural grains (2$-$5\,nm) appears to be irrelevant, we deal with three 
larger scales over distances of tens of nanometers ($L_i\approx 100$\,nm): a) the electronic inhomogeneity of the
pseudogap seen by STS [Fig.\,1(b)]; b) the scale of the 0D AL behavior seen in transport, and c) the topographic 
smooth landscape [Fig.\,1(a)]. Although the scale b) is obtained from transport and cannot be easily connected 
to a spatial structure, it is quite tempting to associate the electronic scales of pseudogap and 0D transport 
to the scale on which Cooper pairs tend to localize (before they eventually condense on the infinite scale of 
the 2D transition). 
On the other hand, the fact that there is no correlation between the topographical map and the superconducting gap map 
proves that the structure is not responsible for the gap inhomogeneities in a trivial way. i.e. it is not some local parameter variation 
(for instance thickness or stoichiometry) that induces a locally-correlated variation of the superconducting properties.  
This does not mean that disorder or structural inhomogeneities are irrelevant. They are most probably relevant, but in a  complex 
manner such as, for instance, disorder is relevant for localization but localization length is not simply the distance between impurities.
 
In any case the main result of our work is that a 0D physics is found to spontaneously emerge in transport 
measurements on the same length scales as the length scale for the superconducting inhomogeneities, 
much larger than the structural crystallites, indicating that this structural complexity has an 
electronic counterpart producing anomalous diffusion of the Cooper pairs.

%%%%%%%%%%%%%%%%%%%%%%%
\vspace{0.5 truecm}
{\bf Conclusion}
%%%%%%%%%%%%%%%%%%%%%%%

The emergence of (glassy) inhomogeneous superconducting phases out of
homogeneously disordered films has been proposed both theoretically \cite{Ioffe:2010dp,Lemarie:2013bk,Bouadim:2011hx} 
and measured experimentally \cite{Sacepe:2011jm,Noat:2013bu,Kamlapure:2013kh}. Our STS measurements have evidenced 
electronic inhomogeneities with typical correlation length ($L_i\approx 100$\,nm) and typical domain size $L_i/2$ 
in the superconducting state below $T_C$ and in the superconducting fluctuations above $T_C$ in NbN ultrathin films. 
These inhomogeneities, which are much larger than the structural grains of the film made of nanometer-sized 
nanocrystals, correspond to simultaneous spatial variations of the energy of the superconducting gap and of the 
energy-integrated in-gap conductance, and are found to become dominating for the thinner samples, where they 
strongly affect the superconductive thermal fluctuations. Specifically, we have shown that, for films with 
nominal thickness $d<2.3$\,nm or $T_C<0.3\,T_C^{bulk}$, these inhomogeneities are associated with specific exponents 
in the dependence of the gaussian superconducting fluctuations with reduced temperature. These exponents 
are consistent with AL fluctuations confined into 0D supergrains of size $l \sim L_i/2$.
On the other hand, at low temperature, the analysis of the magnetic field driven transition is consistent with a 2D 
quantum XY model, which is also evocative of 0D supergrains coupled \textit{\`a la} Josephson. In this regime, the 
Cooper pairs are tightly bound and the transition is ruled by supergrain dephasing, in contrast to the transition 
at finite temperature where AL \textit{amplitude fluctuations} are instead observed.

We propose that, above $T_C$, diffusion slowing down of the Cooper pairs in the supergrains increases their 
lifetime and leads to a depression in the density of states at the Fermi level, yielding a fluctuational pseudogap. 
The origin of these inhomogeneities as well as of this anomalous diffusion process is still to be investigated. In 
any case, this work appeals for a new theoretical microscopic description for this peculiar state of matter, with interplay of localization and pairing,
intermediate between the Fermi and the Bose insulators paradigms.

%%%%%%%%%%%%%%%%%%%%%%%
\vspace{0.5 truecm}
{\bf Methods}
%%%%%%%%%%%%%%%%%%%%%%%

{\em Scanning tunneling spectroscopy}
The electronic inhomogeneities in the superconducting state were probed by using scanning tunneling spectroscopy (STS) 
at 300\,mK or 4.2\,K and establishing full $I(V)$ STS grids of typically 256$\times$256 resolution over a 
300$\times$300 nm$^2$ area. The $dI/dV$ tunneling conductance spectra were obtained by numerical derivation of the 
$I(V)$ data. The conductance maps shown in Fig.\,6 of the Appendix were obtained by integrating in energy the individual 
$dI/dV$ conductance spectra in an energy range of $\pm 0.9$\,meV. The radial profile shown in Fig.~\ref{STS}(f), 
extracted from the autocorrelation map [Fig.~\ref{STS}(d)] of the gap map [Fig.~\ref{STS}(b)], is the circular average 
of all profiles passing through the center of the autocorrelation map. The measurements were performed with a Pt/Ir 
tip. 

{\em Transport measurements}
Transport measurements were carried out by means of a standard four-probe technique in a 14\,T Quantum Design 
PPMS (Physical Properties Measurement System). Resistance data was extracted using either a dc current or an ac 
current at 17\,Hz. Four-probe contacts were made by directly bonding Al wires to the NbN films. Six different 
samples were experimentally analyzed corresponding to different thicknesses: 2.10\,nm ($Y_0$), 2.14\,nm ($X_0$), 
2.16\,nm ($A_2$, or $A_4$, after aging), 2.33\,nm ($B_2$), 2.5\,nm ($C_1$) and 2.9\,nm ($F_0$). The resistivity was measured as a function of temperature with various magnetic field ranging from 0\,T to 14\,T and for sample $A_2$ 
with various applied currents in the vicinity of the QCP.

The critical temperature was determined by using the extrapolation to zero resistance of the tangent 
at the inflection point of the resistance versus temperature curve. See the values in Table 1 of the Appendix.

High magnetic field measurements  were performed at LNCMI (Toulouse) on sample $F_0$, using pulsed fields up to 
55\,T. The ac resistance was measured using four probe technique at a frequency of 50\,kHz. This was used to perform scaling analysis on sample $F_0$ at the vicinity of the QCP.

In order to determine the paraconductance contribution, one needs to extract the normal state conductance.
For samples $A_2$, $A_4$ and $Y_0$, that were close enough to the insulating transition, so that the normal state was
recovered under a perpendicular 14\,T magnetic field, we identified $\sigma_N(T)$ with
$\sigma_{14\,\mathrm{T}}(T)$. For the thicker samples  $B_2$, $C_1$ and $F_0$, the normal state was not recovered 
under 14\,T and we extrapolated the normal state from the high-temperature data (typically between 80K ad 150K) using the same fitting law that 
worked for the 14\,T resistance of the thinner samples, but with different parameters, namely 
$\rho_N(T)=a_3L_T^3+a_1L_T+a_0$, with $L_T\equiv \ln(1/T)$. (For $X_0$, that was measured \textit{in situ} inside the STM cryostat, a similar extrapolation was used for the normal state.) This extrapolation may induce an error on the 
\textit{absolute amplitude} of $\Delta \sigma $, but not on the exponent governing the temperature dependence of 
the divergent term in $\sigma(T)$ at $T=T_C$, since no divergence at $T_C$ is present in the normal state conductance 
nor in its fitting law. 

\section{Acknowledgments}

This work has been supported by a SESAME grant from Region Ile-de-France. Collaboration with Universit\`a di Roma 
La Sapienza was supported by CNRS through a PICS program (S2S)  and by a Joliot grant from ESPCI ParisTech. The work at INSP was supported by the Emergence-UPMC program. We acknowledge the support of the LNCMI-CNRS, member of the European Magnetic Field Laboratory (EMFL).
\newpage

{\bf \centerline{APPENDIX}}

\vspace{1truecm}
\section{Sample fabrication and characterisation}

The NbN films were grown at Karlsruhe Institute of Micro and Nanoelectronic Systems, on the optically polished side of 
sapphire substrates by means of dc reactive magnetron sputtering of a pure Nb target in an $\mathrm{Ar+N_{2}}$ gas mixture 
as described in Ref. \cite{Semenov:2009em}. The deposition 
process was optimized in order to provide the highest transition temperature. 

\begin{figure}
\centering
\includegraphics[width=0.45\textwidth]{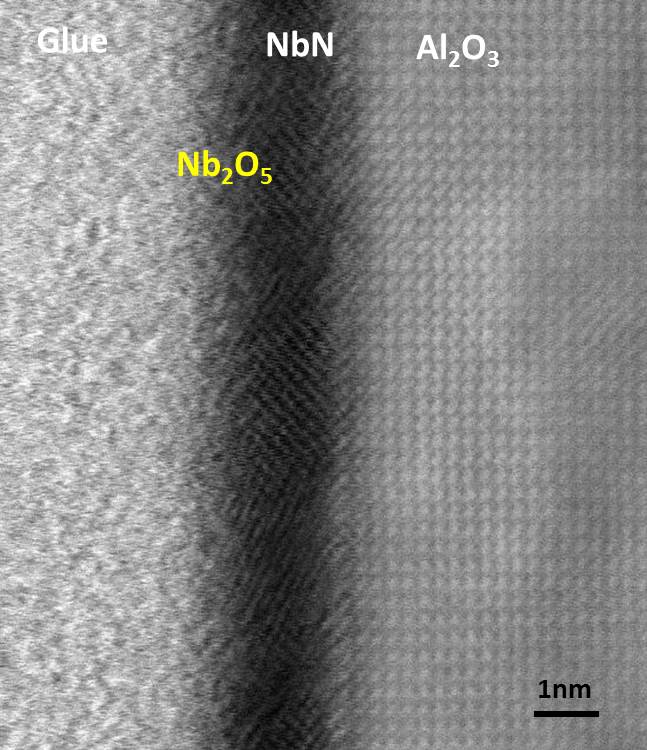}
\caption{High-resolution transmission electron microscopy (HRTEM) image of a 2.33\,nm-thick NbN film grown on sapphire 
substrate. The NbN film consists of nanocrystals of typical size of approximately 2-5\,nm, 
linked together by crystalline boundaries. The presence of a $\simeq$ 1\,nm thick insulating oxide layer (possibly $Nb_2O_5$) on top of the NbN film was inferred from a chemical analysis with scanning TEM measurements across the layers and from AFM measurements and is not visible on this image.}
\label{fig:tem}
\end{figure}

The thickness was evaluated during the deposition of NbN at the Karlsruhe Institute of Micro and Nanoelectronic Systems, 
by time of exposure of the substrate and measurement of the deposition rate. Figure\,\ref{fig:tem} shows a high-resolution 
transmission electron microscopy image of a 2.33\,nm-thick NbN film. The atomic rows of the sapphire substrate and of the NbN layers are clearly visible. The NbN film consists 
of nanocrystals of typical size of approximately 2-5\,nm, linked together by crystalline boundaries.

Both the chemical analysis performed across the film and the atomic force microscopy (AFM) measurements indicated that an oxide 
layer (presumably $\mathrm{Nb_{2}O_{5}}$) is present on top of the NbN films. This layer appears within a few minutes after 
exposing the film to air for the first time. The thickness of the oxide layer on newly prepared films varies in the range  0.5 
to 1\,nm. Further oxidation proceeds much slower. The transition temperature of samples with $T_C \lesssim 10$\,K was actually 
found to decrease with time and thermal cycling. Simultaneously, the normal-state resistance was found to increase. 
This fact was used experimentally to tune the superconductor/insulator transition in sample $A_4$ which corresponds to aged $A_2$. 
We were able to establish that the top oxide layer is insulating and not conducting. The dependency of the STM tunneling current versus tip-sample distance shows an exponential decay whose exponent is directly linked to the effective work function  of the sample. For a metallic sample, it is typically around 4eV. When an ultrathin (ie few atomic monolayers thick) insulating barrier on top of a metal is present, the effective work function is lowered (see \cite{Ploigt:2007ds}), typically by a factor of two at least. We have observed that the dependency of the tunneling current shows such a reduced work function, thus supporting the existence of an insulating oxide layer. Therefore the existence of a top metallic layer such as NbO for instance seems unlikely in our case.
The effect of this oxide layer on the superconducting properties of ultra thin NbN films is discussed in more details in 
\cite{Noat:2013bu}.

\section{ Scanning tunneling spectroscopy (STS) data}

\begin{figure*}
\centering
\includegraphics[width = \textwidth]{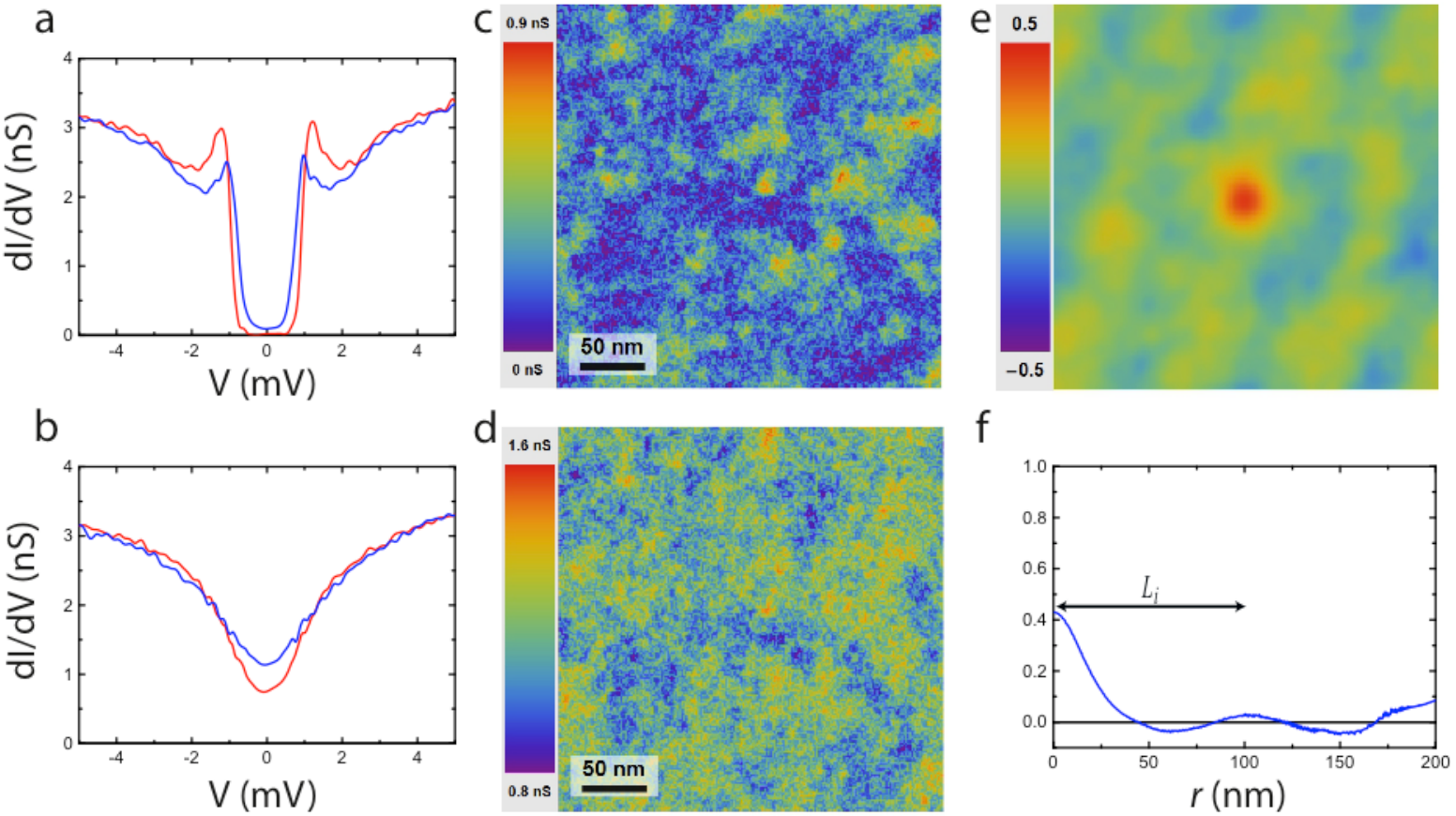}
\caption{$dI/dV$ tunneling conductance spectra as a function of bias voltage measured by STS on sample $X_0$ at a) 300\,mK, b) 
4.2\,K. A spatially varying pseudogap is measured at 4.2\,K, having the characteristic energy scale of the superconducting gap. 
The blue spectra are measured on the lower gap areas (blue areas) of the gap map shown in Fig.\,1b of the main text, while red 
spectra are measured on larger gap areas of Fig.\,1b. c) $dI/dV$ conductance map integrated in the energy range $\pm 0.9$\,meV 
(inside the low-temperature energy gap region) at 300\,mK. The observed inhomogeneities correspond to those seen in the gap map 
of the main text shown in panel Fig.\,1b. d) Same quantity as in c) but at 4.2\,K. This map evidences inhomogenities above $T_C$ 
in the pseudogap features. e) Cross-correlation map between c) and d) showing that the inhomogeneities at 300\,mK and at 4.2\,K 
are strongly correlated. f) Radial profile extracted from the cross-correlation map e). This analysis allows us to conclude that 
the length scale $L_i$ characteristic of the superconducting inhomogeneities is the same at 300\,mK and at 4.2\,K.}
\label{condmap}
\end{figure*}

We present in Fig.\ref{condmap}a) and b) the representative $dI/dV(V)$ tunneling spectra measured at 300\,mK and 4.2\,K on 
the area of interest shown in Fig.\,1a of the main text. The blue spectra are measured on the lower gap areas (blue areas) of 
the gap map presented in Fig.\,1b of the main text, while red spectra are measured on larger gap areas of Fig.\,1b. 
It is seen 
that a spatially varying pseudogap is measured at 4.2\,K, having the characteristic energy scale of the superconducting gap 
measured at low temperature. Figs.\,\ref{condmap}c) and d) present $dI/dV$ conductance maps integrated in the energy range of 
the low-temperature superconducting gaps, at 300\,mK (c) and at 4.2\,K (d). It is clear from this figure that both 
superconducting correlations and superconducting inhomogeneities persist above $T_C$. In addition, our analysis shows that 
these two conductance maps c) and d) measured over the same topographic area are strongly correlated. This result can be seen 
in Fig. \ref{condmap}e) where the cross-correlation map between \ref{condmap}c) and \ref{condmap}d) is plotted. This proves that 
the inhomogeneities in the energy gap value below $T_C$ and in the pseudogap features above $T_C$ are strongly 
spatially correlated. Furthermore, the radial profile shown in Fig. \ref{condmap}f), extracted from a circular average of the cross-correlation 
map e), allows us to infer that the characteristic correlation length $L_i$ of these inhomogeneities at 4.2\,K is comparable to the 
one seen at 300\,mK, of size about 100\,nm.
We have also performed STS measurements on sample $X_0$ at higher temperature so well in the 0D regime (around 7K) which show similar results as the one presented in Fig. \ref{condmap}b) and Fig. \ref{condmap}d) at 4.2K. The only difference between the 7K and 4.2K data is that the spectroscopic features related to the pseudogap are smaller in amplitude at 7K. But the regions where the pseudogap is still present (corresponding to the supergrains) have the same characteristic size and spacing as at 300mK or at 4.2K. 

\section{Analysis of the paraconductivity: Fitting parameters for the normal state resistance}

For samples $A_2$, $A_4$ and $Y_0$, that were close enough to the insulating transition, so that the normal state was
recovered under a perpendicular 14\,T magnetic field, we identified $\sigma_N(T)$ with
$\sigma_{14\,\mathrm{T}}(T)$. 
The resistance of sample $X_0$ was measured {\it in situ} inside the STM cryostat up to 30\,K, therefore the measure under 14T magnetic field could not be performed, and we extrapolated the normal state.
For the thicker samples, $B_2$, $C_1$ and $F_0$, superconductivity could not be completely suppressed by
a magnetic field of $14$\,T. In such cases, the normal-state resistance was fitted by an expression that worked 
well for the thinner samples under 14\,T, 
\be
\rho_N(T)=a_3L_T^3+a_1L_T+a_0,  \label{fitetnor}
\ee

with $L_T\equiv \ln(1/T)$. We determined 
the parameters $a_{0,1,3}$ fitting the high-temperature resistance, typically in the range 80$-$150\,K
(except for the sample $X_0$, for which the range was forcedly smaller, 10$-$30\,K). The 
fitting parameters are found in Tab. \ref{table-fit}.
\begin{table}
\centering
\begin{tabular}{||c|c|c|c|c|c||}
\hline
sample name & $\displaystyle{\mathrm{thickness\atop (nm)}}$  &
$\displaystyle{T_C\atop\mathrm{(K)}}$ & $\displaystyle{R_{sq}(300\,\mathrm{K})\atop(\mathrm{k\Omega/sq})}$ &
$\displaystyle{T_C^{2D}\atop\mathrm{(K)}}$ & $\displaystyle{T_C^{0D}\atop\mathrm{(K)}}$ \\
\hline
$F_{0}$ & 2.90  & 9.0 & 1390 & 8.6 & - \\
\hline
$C_{1}$ & 2.50 & 9.4 & 2370 & 9.5 & - \\
\hline
$B_{2}$ & 2.33 & 7.1 & 2450 & 7.1 & - \\
\hline
$A_{2}$ & 2.16 & 4.5 & 3150 & 4.9 & 4.4\\
\hline
$A_{4}$  (aged $A_2$) & 2.16 & 2.4 & 4570 & 3.0 & 2.9\\
\hline
$X_{0} $ & 2.14 & 3.8 & 3800 & 3.9 & 3.6\\
\hline
$Y_{0}$ & 2.10 & 4.3  & 2350 & 4.9 & 4.3\\
\hline
\end{tabular}
\caption{Sample nominal thickness, critical temperature $T_C$ (defined as the extrapolation to zero resistance of the tangent 
at the inflection point of the resistance curve), resistance per square $R_{sq}$ at room temperature for the different 
samples studied, and ``fluctuational'' critical temperature for the 2D ($T_C^{2D}$) and 0D ($T_C^{0D}$) AL regime.}
\label{tableau}
\end{table}

\begin{table}
\centering
\begin{tabular}{||c|c|c|c||}
\hline
sample name & $a_0$ $(\mathrm{k\Omega/sq})$ & $a_1$ $(\mathrm{k\Omega/sq})$ & $a_3$ $(\mathrm{k\Omega/sq})$ \\
\hline
$X_{0}$ & 9299 & 1038 & 1.370 \\
\hline
$B_{2}$ & 5804 & 275.3 & 10.53 \\
\hline
$C_{1}$ & 3529 & 32.32 & 6.094 \\
\hline
$F_{0}$ & 2212  & 4.512 & 4.381 \\
\hline
\end{tabular}
\caption{Fitting parameters of the normal state resistance according to the expression 
$\rho_N(T)=a_3L_T^3+a_2L_T^2+a_1L_T+a_0$, with $L_T\equiv \ln(1/T)$. Since the inclusion 
of the $L_T^2$ term did not improve significantly the fits, to reduce the number of fitting parameters, we fixed $a_2=0$.}
\label{table-fit}
\end{table}

\section{Analysis of the paraconductivity: possible observation of a 0d-2d crossover}

\begin{figure}
\centering
\includegraphics[width = 0.45\textwidth]{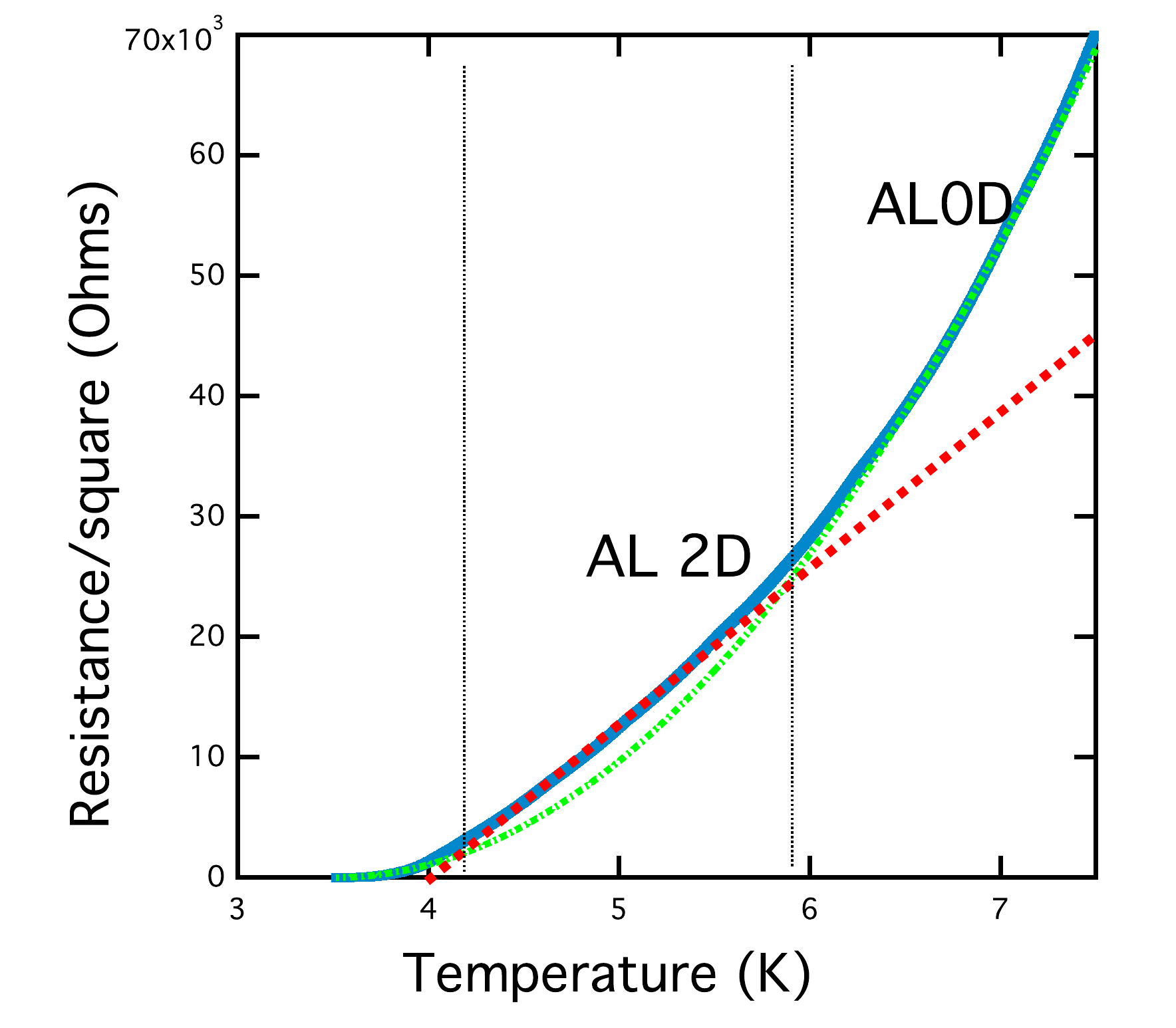}
\caption{Pararesistance $\Delta\sigma^{-1}$ for sample $X_0$ as a function of temperature, illustrating the 
0D AL behavior, where $\Delta\sigma^{-1}\propto (T-T_C^{0D})^2$, and the
crossover to the 2D AL regime close enough to $T_C$, where $\Delta\sigma^{-1}\propto T-T_C^{2D}$. We point
out that the ``fluctuational'' critical temperatures $T_C^{0D}$ and $T_C^{2D}$ need not be the same,
as they enter the expressions of fluctuation contributions living on different length scales.}
\label{aa}
\end{figure}

In Fig.\,\ref{aa} we plot the pararesistance $\Delta\sigma^{-1}$ as a function of temperature 
(and not reduced temperature $\epsilon$) for sample $X_0$. Such plot (similar the right inset of Fig. 3b in the Article for sample $A_2$) allows to visualize both the 
0D AL (quadratic) regime and the 2D (linear) regime closer to $T_C$. Since the two regimes are characterized by 
different ``fluctuational'' critical temperatures, they cannot be both visualized with a single choice of
$T_C$ in the reduced temperature $\epsilon$. Such 0D-2D crossover was indeed observed in all thinner samples $A_2$, $A_4$, $X_0$ and $Y_0$. 

\section{Study of the quantum critical region}

\begin{figure}
\centering
\includegraphics[width = 0.45\textwidth]{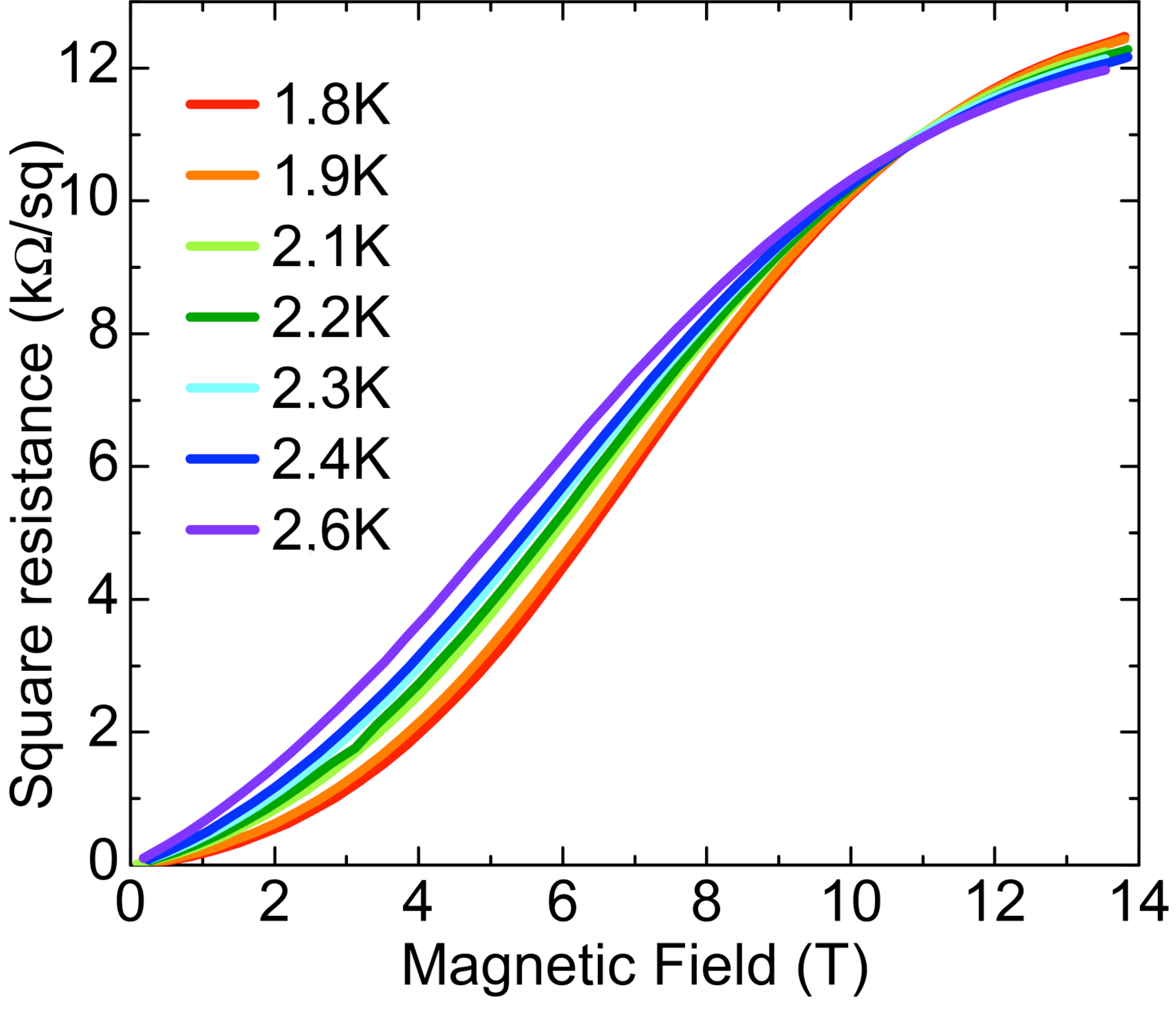}
\caption{Resistance versus magnetic field data for temperature between 1.8\,K (lower curve at low fields) and 2.6\,K 
(upper curve at low fields).}
\label{RH}
\end{figure}

In Fig.\,\ref{RH} are displayed the isothermal resistance versus magnetic field curves for different temperatures. The curves 
for temperatures between 1.8\,K and 2.6\,K exhibit a common crossing point, at $H_C= 10.7$\,T and $R_C=11$\,k$\Omega$, 
corresponding to the quantum critical point. For these measurements, the current and voltage values across the sample were very low 
and the curves were found to be insensitive to amplitude variations of the current bias.
\begin{figure}
\centering
\includegraphics[width = 0.45\textwidth]{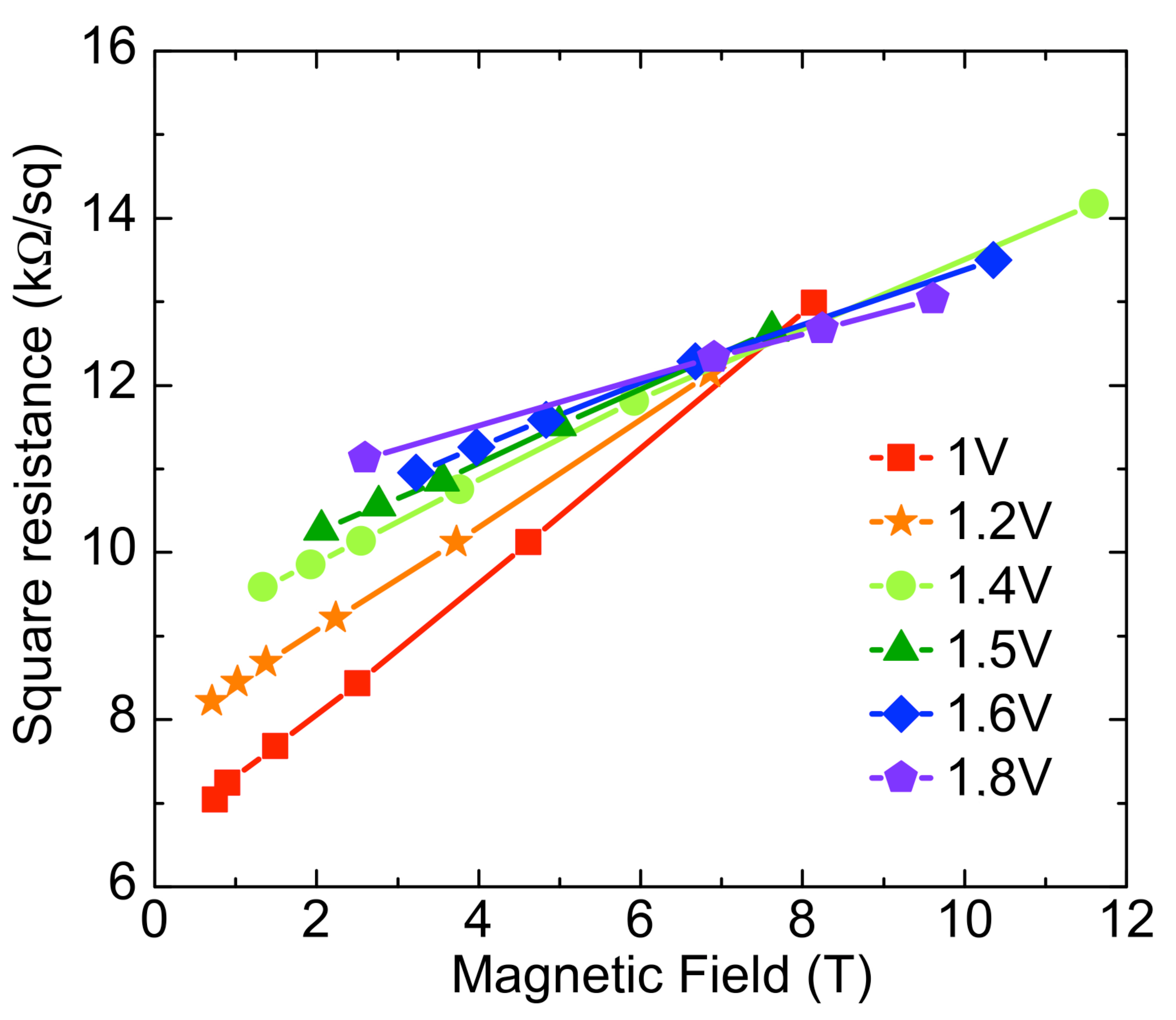}
\caption{Resistance as a function of magnetic field at different electrical fields across the sample at a temperature of 1.9\,K.}
\label{RHE}
\end{figure}
In Fig.\,\ref{RHE} are displayed the isopotential resistance versus magnetic field curves for different values of the voltage 
applied across the sample, corresponding to different values of the electric field. The measurements were performed at 1.9\,K. 
As for the isothermal curves, the isopotential curves are found to cross in a single point. These curves were then used for scaling analysis as described in the Article. 

\section{Theoretical analysis of paraconductivity}

\subsection{Zero dimensional Aslamazov-Larkin (AL) fluctuations}

To discuss AL paraconductivity, we assume overall isotropy (at least in the average) and start from the expression 
\cite{Aslamasov:1968vx,Caprara:2005im}
\be
\delta\sigma_D(\epsilon)=\frac{\pi e^2}{4\hbar}\int\frac{\mathrm d^D \bf q}{(2\pi)^D}\frac{1}{D}\left(
\vec\nabla_{\bf q}\gamma_{\bf q}\right)^2\frac{1}{(\epsilon+\gamma_{\bf q})^3}  
\ee
that reproduces the standard AL result in $D$ dimensions with $\gamma_{\bf q}\equiv\xi_0^2|{\bf q}|^2\equiv\xi_0^2 q^2$ being 
the relaxation rate for each mode $\bf q$. $\xi_0$ is the coherence length and $\epsilon\equiv\log(T/T_C)$, $T_C$ being 
the critical temperature seen by the fluctuating Cooper pairs (which may differ from the actual critical temperature, 
in the presence of crossover phenomena). The above formula is rather general \cite{Caprara:2009dq}, although the 
expression of $\epsilon$ may change depending on the microscopic theory, and the expression 
$\epsilon\equiv\alpha(T-T_C)/T_C$, with a dimensionless prefactor $\alpha$ not necessarily equal to one, might be used. 
However, for definiteness, we adopt henceforth the standard definition of $\epsilon$.

Since $\gamma_{\bf q}$ only depends on $q$, the angular integration factors out and one 
obtains
\be
\delta\sigma_D=\frac{\pi e^2 \Omega_D}{4\hbar D(2\pi)^D}\int\mathrm dq\,q^{D-1}\left(
\frac{\mathrm d\gamma_q}{\mathrm dq}\right)^2\frac{1}{(\epsilon+\gamma_q)^3}, 
\ee
where $\Omega_D\equiv 2\pi^{D/2}/\Gamma(D/2)$ is the surface of the unitary sphere in 
$D$ dimensions and $\Gamma(z)$ is the Euler gamma function. Exploiting the expression of $\gamma_q$ one obtains
\be
\delta\sigma_D=\frac{\pi e^2 \Omega_D\xi_0^4}{\hbar D(2\pi)^D}\int_{0}^{\infty}\mathrm dq\,
\frac{q^{D+1}}{\left(\epsilon+\xi_0^2 q^2\right)^3}.
\ee
When $D\to 0$, $\Omega_D$ inherits the singularity of $\Gamma(D/2)\approx 2/D$, so that $\Omega_D\approx D$,
and
\be
\delta\sigma_{D\to 0}=\frac{\pi e^2 \xi_0^2}{4\hbar\epsilon^2}, \label{eq4}
\ee
i.e., there exists a well defined limit of the AL paraconductivity as $D\to 0$, that scales with $\epsilon^{-2}$, and 
qualitatively agrees with the behavior observed in the thinnest NbN films. This result is converted into the measured 
conductance per square, dividing it by the square of a suitable length scale $l$, related with the size of the fluctuating 
(nano)domains,
\be
\delta\sigma_{\mathrm{measured}}=\left(\frac{\xi_0}{l}\right)^2\frac{\pi e^2}{4\hbar\epsilon^2}. \label{eq5}
\ee
However, although the supergrains may behave as quantum dots, closer to $T_C$ the system is two-dimensional, the 
supergrains are coupled, and the paraconductive fluctuations must cross over to $D=2$. The formal limit does not 
allow for a description of this crossover and the identification of the relevant length scales. A crucial remark 
of Ref. \cite{Caprara:2005im} is that the AL theory can be cast in an even more general form, introducing
a suitable weighted density of states,
\be
\mathcal N_D(\gamma)\equiv \int\frac{\mathrm d^D \bf q}{(2\pi)^D}\left(
\vec\nabla_{\bf q}\gamma_{\bf q}\right)^2\delta(\gamma-\gamma_{\bf q}), \label{eq6}
\ee
which allows to write the paraconductivity in the form
\be
\delta\sigma_D=\frac{\pi e^2}{4\hbar D}\int_{0}^{\infty} \mathrm d\gamma \,\frac{\mathcal N_D(\gamma)}{(\epsilon+\gamma)^3}. \label{eq7}
\ee
The standard AL result is recovered with 
\be
\mathcal N_D(\gamma)\equiv \frac{4\xi_0^4\Omega_D}{(2\pi)^D}
\int_{0}^{\infty}\mathrm dq\, q^{D+1}\delta(\gamma-\xi_0^2 q^2), \label{eq8}
\ee
yielding
\be
\delta\sigma_D=\frac{\pi e^2 \xi_0^{2-D}\Omega_D}{2\hbar D(2\pi)^D}
\int_{0}^{\infty} \mathrm d\gamma \,\frac{\gamma^{D/2}}{(\epsilon+\gamma)^3}. \label{eq9}
\ee

We argue that the observed 0D-2D crossover stems from a different form of 
$\mathcal N_D(\gamma)$, that results from the supergrains (endowed with an internal structure) being 
connected in a 2D network. In the following we thus set $D=2$. The standard 2D AL result 
is found for $\mathcal N_2(\gamma)=\gamma/\pi$, corresponding to diffusion of fluctuating Cooper pairs in two dimensions. 
If instead $\mathcal N_2(\gamma)=\gamma/\pi$, for $\gamma\le\bar\gamma$, and $\mathcal N_2(\gamma)=\bar\gamma/\pi$,
for $\gamma>\bar\gamma$, which describes a slowing-down of the diffusion of fluctuating Cooper pairs above a threshold 
$\bar\gamma$ (e.g., corresponding to trapping of the fluctuations within the supergrains), the paraconductivity is 
\be
\delta\sigma=\frac{e^2}{16\hbar}\frac{\bar\gamma}{\epsilon(\epsilon+\bar\gamma)}, \label{eq10}
\ee

which recovers the 2D behavior for $\epsilon \ll \bar\gamma$, while for $\epsilon\gg\bar\gamma$
\be
\delta\sigma\approx\frac{e^2\bar\gamma}{16\hbar\epsilon^2}, \label{eq11}
\ee

mimicking the 0D behavior at higher temperature. A comparison with the formal extrapolation of AL fluctuations for 
$D\to 0$ yields $\bar\gamma=4\pi(\xi_0/l)^2$. As a matter of 
fact, the 0D-2D crossover is clearly visible for sample $X_0$ ($d=2.14$\,nm) for which a much greater number of points 
was taken in the vicinity of the transition, as well as for the sample $A_2$ ($d=2.16$\,nm), but also for $A_4$ and $Y_0$. Eventually, at temperatures below 
the 2D AL regime, the system may be governed by percolation physics or, alternatively, 
by Berezinski-Kosterlitz-Thouless behavior. 

A rough estimate gives $\epsilon\approx\xi_0^2/\xi^2$, where $\xi$ is the 
temperature dependent correlation length that diverges
at the transition. Then, the 0D-2D crossover takes place at a temperature $\bar T$ such that
$\xi(\bar T)\approx \xi_0\sqrt{\bar\gamma}$. A comparison with the formal 0D limit of the AL theory
allows us to identify $\bar\gamma\approx 4\pi\left(\xi_0/l\right)^2$, hence the size of the
supergrains $l\approx \xi_0\sqrt{4\pi/\bar\gamma}$.

%\bibliography{biblioNbN2}

\newpage

\begin{thebibliography}{10}

\bibitem{Goltsman:2001ea}
Gol'tsman G.N., Okunev O., Chulkova G., Lipatov A., Semenov A., Smirnov K.,
 et al.
\newblock {Picosecond superconducting single-photon optical detector}.
\newblock {\em Applied Physics Letters}, \textbf{79}, 705 (2001).

\bibitem{Hofherr:2010iy}
Hofherr M., Rall D., Ilin K., Siegel M., Semenov A.,  H{\"u}bers H.W., et al.
\newblock {Intrinsic detection efficiency of superconducting nanowire
  single-photon detectors with different thicknesses}.
\newblock {\em Journal of Applied Physics}, \textbf{108}, 014507 (2010).

\bibitem{Goldman:1998jy}
Goldman A.M. and Markovi{\'c} N.
\newblock {Superconductor-Insulator Transitions in the Two-Dimensional Limit}.
\newblock {\em Physics Today}, \textbf{51}, 39 (1998).

\bibitem{Beloborodov:2007hq}
Beloborodov I., Lopatin A., Vinokur V., and Efetov K.
\newblock {Granular electronic systems}.
\newblock {\em Reviews of Modern Physics}, \textbf{79}, 469-518 (2007).

\bibitem{FinkelShtein:1987tl}
Finkel'Shtein A.M.
\newblock {Superconductivity-transition temperature in amorphous films}.
\newblock {\em Pisma v Zhurnal Eksperimentalnoi i Teoreticheskoi Fiziki},
\textbf{  45}, 37-40, (1987).

\bibitem{VallesJr:1992va}
Valles J.M., Jr, Dynes R.C., and  Garno J.R.
\newblock {Electron tunneling determination of the order-parameter amplitude at
  the superconductor-insulator transition in 2D}.
\newblock {\em Physical Review Letters},\textbf{ 69}, 3567-3570 (1992).

\bibitem{Fisher:1990zza}
Fisher M.P.A, Grinstein G., and Girvin S.M.
\newblock {Presence of quantum diffusion in two dimensions: Universal
  resistance at the superconductor-insulator transition}.
\newblock {\em Physical Review Letters},\textbf{ 64}, 587-590, (1990).

\bibitem{Fisher:1990zz}
Fisher M.P.A.
\newblock {Quantum phase transitions in disordered two-dimensional
  superconductors}.
\newblock {\em Physical Review Letters}, \textbf{65}, 923-926(1990).

\bibitem{castellani:2012}
Seibold G., Benfatto L., Castellani C., and Lorenzana J.
\newblock {Superfluid Density and Phase Relaxation in Superconductors with Strong Disorder}.
\newblock {\em Physical Review Letters}, \textbf{108}, 207004 (2012).

\bibitem{Feigelman:2007bq}
M~Feigel'man, L~Ioffe, V~Kravtsov, and E~Yuzbashyan.
\newblock {Eigenfunction Fractality and Pseudogap State near the
  Superconductor-Insulator Transition}.
\newblock {\em Physical Review Letters},\textbf{ 98}, 027001(2007).

\bibitem{Feigelman:2010hp}
Feigel'man M.V.,  Ioffe L.B., Kravtsov V.E., and Cuevas E.
\newblock {Annals of Physics}.
\newblock {\em YAPHY}, \textbf{325},1390-1478 ( 2010).

\bibitem{Ioffe:2010dp}
L~B Ioffe and Marc M{\'e}zard.
\newblock {Disorder-Driven Quantum Phase Transitions in Superconductors and
  Magnets}.
\newblock {\em Physical Review Letters}, \textbf{105}, 037001 (2010).

\bibitem{Hollen:2013ht}
Hollen S.M., Fernandes G.E., Xu J.M., and Vall{\`e}s J.M.
\newblock {Collapse of the Cooper pair phase coherence length at a
  superconductor-to-insulator transition}.
\newblock {\em Physical Review B}, \textbf{87},054512(2013).

\bibitem{Stewart:2007uz}
Stewart M.D., Yin A., Xu J.M., and Valles J.M.
\newblock {Superconducting pair correlations in an amorphous insulating
  nanohoneycomb film}.
\newblock {\em Science}, \textbf{318}, 1273-1275 (2007).

\bibitem{Kopnov:2012bx}
Kopnov G., Cohen O., Ovadia M., Lee K.H., Wong C.C., and Shahar D.
\newblock {Little-Parks Oscillations in an Insulator}.
\newblock {\em Physical Review Letters}, \textbf{109}, 167002 (2012).

\bibitem{Lemarie:2013bk}
Lemari{\'e} G., Kamlapure A., Bucheli D., Benfatto L., Lorenzana J., Seibold G., et al. 
\newblock {Universal scaling of the order-parameter distribution in strongly
  disordered superconductors}.
\newblock {\em Physical Review B}, \textbf{87}, 184509 ( 2013).

\bibitem{Sacepe:2008jx}
Sac{\'e}p{\'e} B., Chapelier C., Baturina T.I., Vinokur V.I. , Baklanov M.R., and Sanquer M.
\newblock {Disorder-Induced Inhomogeneities of the Superconducting State Close
  to the Superconductor-Insulator Transition}.
\newblock {\em Physical Review Letters}, \textbf{101}, 157006 (2008).

\bibitem{Sacepe:2011jm}
Sac{\'e}p{\'e} B., Dubouchet T., Chapelier C., Sanquer M.,
  Ovadia M., Shahar D., et al.
\newblock {Localization of preformed Cooper pairs in disordered
  superconductors}.
\newblock {\em Nature Physics}, \textbf{7}, 239-244 (2011).

\bibitem{Bouadim:2011hx}
Bouadim K., Loh Y.L., Randeria M. and Trivedi N.
\newblock {Single- and two-particle energy gaps across the disorder-driven
  superconductor--insulator transition}.
\newblock {\em Nature Physics}, \textbf{7}, 884-889 (2011).

\bibitem{Trivedi:2012cj}
Trivedi N., Loh Y.L., Bouadim K., and Randeria M.
\newblock {Emergent granularity and pseudogap near the superconductor-insulator
  transition}.
\newblock {\em Journal of Physics: Conference Series}, \textbf{376}, 012001 ( 2012).

\bibitem{Kamlapure:2013kh}
Kamlapure A., Das T., Ganguli S.C., Parmar J.B., Bhattacharyya S., and Raychaudhuri P.
\newblock {Emergence of nanoscale inhomogeneity in the superconducting state of
  a homogeneously disordered conventional superconductor}.
\newblock {\em Scientific Reports}, \textbf{3} , 2979 (2013).

\bibitem{Noat:2013bu}
Noat Y., Cherkez V., Brun C, Cren T., Carbillet C., Debontridder F., et al.
\newblock {Unconventional superconductivity in ultrathin superconducting NbN
  films studied by scanning tunneling spectroscopy}.
\newblock {\em Physical Review B}, \textbf{88}, 014503 (2013).

\bibitem{Sacepe:2010gt}
Sac{\'e}p{\'e} B., Chapelier C., Baturina T.I., Vinokur V.I. , Baklanov M.R., and Sanquer M.
\newblock {Pseudogap in a thin film of a conventional superconductor}.
\newblock {\em Nature Communications}, \textbf{1},140 (2010).

\bibitem{Abrahams:1970tn}
Abrahams E., Redi M., and Woo J.
\newblock {Effect of Fluctuations on Electronic Properties above the
  Superconducting Transition}.
\newblock {\em Physical Review B}, \textbf{1}, 208-213 (1970).

\bibitem{Semenov:2009em}
Semenov A., G{\"u}nther B., B{\"o}ttger U.,  H{\"u}bers H.W., Bartolf H., Engel A.,
  et al.
\newblock {Optical and transport properties of ultrathin NbN films and
  nanostructures}.
\newblock {\em Physical Review B}, \textbf{80}, 054510 (2009).

\bibitem{Aslamasov:1968vx}
Aslamasov L.G. and Larkin A.I.
\newblock {The influence of fluctuation pairing of electrons on the
  conductivity of normal metal}.
\newblock {\em Physics Letters A}, \textbf{26}, 238-239(1968).

\bibitem{Maki:1968vo}
Maki K.
\newblock {Critical fluctuation of the order parameter in a superconductor. I}.
\newblock {\em Progress of Theoretical Physics}, \textbf{40}, 193-200 (1968).

\bibitem{Thompson:1970ue}
Thompson R.S.
\newblock {Microwave, flux flow, and fluctuation resistance of dirty type-II
  superconductors}.
\newblock {\em Physical Review B}, \textbf{1}, 327(1970).

\bibitem{Beloborodov:1999va}
Beloborodov I.S. and Efetov K.B.
\newblock {Negative Magnetoresistance of Granular Metals in a Strong Magnetic
  Field}.
\newblock {\em Physical Review Letters}, \textbf{82}, 3332-3335 (1999).

\bibitem{Beloborodov:2000vw}
Beloborodov I.S., Efetov K.B., and Larkin A.I.
\newblock {Magnetoresistance of granular superconducting metals in a strong
  magnetic field}.
\newblock {\em Physical Review B}, \textbf{61}, 9145-9161 (2000).

\bibitem{Kirtley:1974tb}
Kirtley J., Imry Y., and Hansma P.K.
\newblock {Fluctuation-Induced Conductivity Above the Critical Temperature in
  Small-Particle Arrays}.
\newblock {\em Journal of Low Temperature Physics}, \textbf{17}, 247 (1974).

\bibitem{Deutscher:1974wn}
Deutscher G. , Imry Y., and Gunther L.
\newblock {PhysRevB.10).4598}.
\newblock {\em Physical Review B}, \textbf{10}, 4598 (1974).

\bibitem{Civiak:1976uh}
Civiak R.I., Elbaum C., Nichols L.F., Kao H.I., and  Labes M.M.
\newblock {Fluctuation-Induced Conductivity and Dimensionality in Polysulfur
  Nitride}.
\newblock {\em Physical Review B},\textbf{ 14}, 5413-5421 (1976).

\bibitem{WOLF:1977wl}
Wolf S. and Lowrey W.H.
\newblock {Zero Dimensionality and Josephson Coupling in Granular Niobium
  Nitride}.
\newblock {\em Physical Review Letters}, \textbf{39}, 1038-1041 (1977).

\bibitem{Belevtsev:1983wd}
Belevtsev B.I. and Komnik Y.F.
\newblock {Superconducting fluctuations above T$_{c}$ in quench-condensed
  hydrogen-doped indium films: transition from two- to zero-dimensional}.
\newblock {\em Fis. Nisk. Temp.},\textbf{ 9}, 581(1983).

\bibitem{Reizer:1992wr}
Reizer M.Y.
\newblock {Fluctuation conductivity above the superconducting transition:
  Regularization of the Maki-Thompson term}.
\newblock {\em Physical Review B}, \textbf{45},12949 (1992).

\bibitem{Sondhi:1997wz}
Sondhi S.L. , Girvin S.M., Carini J.P., and Shahar D.
\newblock {Continuous quantum phase transitions}.
\newblock {\em Reviews of Modern Physics}, \textbf{69}, 1-19(2012).

\bibitem{Yazdani:1995ut}
Yazdani A. and Kapitulnik A.
\newblock {Superconducting-insulating transition in two-dimensional a-MoGe thin
  films}.
\newblock {\em Physical Review Letters},\textbf{ 74}, 3037-3040 (1995).

\bibitem{Markovic:1999vy}
Markovi{\'c} N., Christiansen C., Mack A.M., Huber W.H., and Goldman A.M.
\newblock {Superconductor-insulator transition in two dimensions}.
\newblock {\em Physical Review B}, \textbf{60}, 4320 (1999).

\bibitem{Lawrence:1971}
Lawrence W.E. and Doniach S.
\newblock {Theory of layer structure superconductors}.
\newblock {\em Proc. 12th Int. Conf. on Low Temp. Phys.}, 361 (1971).

\bibitem{Halperin:1979to}
Halperin B.I. and Nelson D.R.
\newblock {Resistive Transition in Superconducting Films}.
\newblock {\em Journal of Low Temperature Physics},\textbf{ 36}, 599-616 (1979).

\bibitem{Herbut:2001hn}
Herbut I.
\newblock {Quantum Critical Points with the Coulomb Interaction and the
  Dynamical Exponent: When and Why z=1}.
\newblock {\em Physical Review Letters}, \textbf{87},137004 (2001).

\bibitem{Kisker:1997un}
Kisker J. and Rieger H.
\newblock {Bose-glass and Mott-insulator phase in the disordered boson Hubbard
  model}.
\newblock {\em Physical Review B}, \textbf{55}, R11981 (1997).

\bibitem{Aubin:2006ju}
Aubin H., Marrache-Kikuchi C., Pourret A., Behnia K., Berg{\'e} L., Dumoulin L., et al.
\newblock {Magnetic-field-induced quantum superconductor-insulator transition
  in $Nb_{0.15}Si_{0.85}$}.
\newblock {\em Physical Review B}, \textbf{73}, 094521 (2006).

\bibitem{Hebard:1990up}
Hebard A.F. and Paalanen M.A.
\newblock {Magnetic-field-tuned superconductor-insulator transition in
  two-dimensional films}.
\newblock {\em Physical Review Letters}, \textbf{65}, 927-930 (1990).

\bibitem{Kapitulnik:2001cu}
Kapitulnik A., Mason N., Kivelson S., and Chakravarty S.
\newblock {Effects of dissipation on quantum phase transitions}.
\newblock {\em Physical Review B}, \textbf{63}, 125322 (2001).

\bibitem{Steiner:2005ks}
Steiner M. and Kapitulnik A.
\newblock {Superconductivity in the insulating phase above the field-tuned
  superconductor--insulator transition in disordered indium oxide films}.
\newblock {\em Physica C: Superconductivity}, \textbf{422}, 16-26 (2005).

\bibitem{Ploigt:2007ds}
Hans-Christoph Ploigt, Christophe Brun, Marina Pivetta, François Patthey, and Wolf-Dieter Schneider
\newblock{Local work function changes determined by field emission resonances : NaCl?Ag(100)}. 
\newblock{\em Phys. Rev. B},  \textbf{76}, 195404  (2007).

\bibitem{Caprara:2005im}
S Caprara, M Grilli, B Leridon, and J Lesueur. 
\newblock{Extended paraconductivity regime in underdoped cuprates}.\newblock{\em Physical Review B}, \textbf{72}
\newblock{104509 (2005)}.

\bibitem{Caprara:2009dq}S Caprara, M Grilli, B Leridon, and J Vanacken. 
\newblock{Paraconductivity in layered cuprates behaves as if due to pairing of nearly free quasiparticles}. 
\newblock{\em Physical Review B}, \textbf{79}, 024506  (2009).

\end{thebibliography}
%\begin{thebibliography}{99}
%\bibitem{AL} L. G. Aslamazov and A. I. Larkin, Phys. Lett. {\bf 26} A, 238 (1968);
%Sov. Phys. Solid State {\bf 10}, 875 (1968).
%\bibitem{noi} S. Caprara, M. Grilli, B. Leridon, and J. Lesueur, Phys. Rev. B {\bf 72}, 104509 (2005).
%\bibitem{cg} S. Caprara, M. Grilli, B. Leridon, and J. Vanacken, Phys. Rev. B {\bf 79}, 024506 (2009).
%\end{thebibliography}

\vfil
\eject

\section{Author contribution}

The samples were made by KI. The TEM  experiments were performed by DD and LL. The transport measurements were proposed 
by BL and carried out by CC. The high magnetic field measurement were proposed by BL and carried out by CC, WT and BV. 
The STS measurement were proposed by DR, TC and CB and carried out by CC, CB, TC, FD and DR. The analysis of the 
transport data was done by CC, BL, SC and MG and of tunneling data by CC, TC and CB. The 0D AL paraconductivity 
was proposed by BL and calculated by SC and MG. The paper was written by BL, SC, MG and CB. All authors discussed 
the results and commented on the manuscript.\\

\end{document}